\documentclass[a4paper]{jpconf}
\usepackage{graphicx}

\def\gtsima{$\; \buildrel > \over \sim \;$}
\def\ltsima{$\; \buildrel < \over \sim \;$}
\def\gtrsim{\lower.5ex\hbox{\gtsima}}
\def\lesssim{\lower.5ex\hbox{\ltsima}}
\def\apj{Astrophysical Journal}
\def\apjl{\apj Letter}
\def\nat{Nature}
\def\mnras{MNRAS}
\def\prd{Physical Review Documents}
\def\aap{Astronomy \& Astrophysics}
\def\pasa{Publications of the Astronomical Society of Australia}

\def\aapr{The Astronomy and Astrophysics Review}
\def\araa{Annual Review of Astronomy and Astrophysics}
\def\aj{Astronomical Journal}

\def\planss{Planetary and Space Science}
\def\pasj{Publications of the Astronomical Society of Japan}

\begin{document}
%\title{Astrophysics of black hole binaries:
%what is new after four detections?}

\title{Black hole demography at the dawn of gravitational-wave astronomy:\\
state of the art and future perspectives}

\author{Michela Mapelli$^{1,2}$}

\address{$^1$ INAF -- Osservatorio Astronomico di Padova, Vicolo dell'Osservatorio 5, I--35122, Padova, Italy}

\address{$^2$ Institut f\"ur Astro- und Teilchenphysik, Universit\"at Innsbruck, Technikerstrasse 25/8, A--6020 Innsbruck, Austria}

\ead{michela.mapelli@oapd.inaf.it}

\begin{abstract}
The first LIGO-Virgo detections have confirmed the existence of massive black holes (BHs), with mass $30-40$ M$_\odot$. Such BHs might originate from massive metal-poor stars ($Z<0.3$ Z$_\odot$) or from gravitational instabilities in the early Universe. The formation channels of merging BHs are still poorly constrained. The measure of mass, spin and redshift distribution of merging BHs will give us fundamental clues to distinguish between different models. Also, a better understanding of several astrophysical processes (e.g. common envelope, core-collapse supernovae, and dynamical evolution of BHs) is decisive, to shed light on the formation channels of merging BHs. 
\end{abstract}

%\section{What we know from four detections?}
\section{Lesson learned from the first gravitational wave detections}
Five detections of black hole (BH) mergers have been reported so far by the LIGO-Virgo collaboration and several tens of new detections are expected in the next few years \cite{abbott2016a,abbott2016b,abbott2016c,abbott2016d,abbott2017a,abbott2017b,abbott2017c}: the era of gravitational wave astronomy has just begun. %The era of gravitational wave astronomy has just begun, with crucial implications for the astrophysics of BHs: the existence of BH binaries is confirmed, and we also know that such systems can host massive BHs, i.e. BHs with mass higher than $\sim{}20$ M$_\odot$. 
% This result opens the era of gravitational wave astronomy and has crucial implications for the astrophysics of BHs: it confirms the existence of BH binaries, and it tells us that such systems can host massive BHs, i.e. BHs with mass higher than $\sim{}20$ M$_\odot$. 

 The first detections already have crucial implications for the astrophysics of BHs: they confirm the existence of BH binaries and tell us  that such systems can host massive BHs, i.e. BHs with mass higher than $\sim{}20$ M$_\odot$. 

This was a big surprise for the astrophysics community, considering that the only BHs for which we had a dynamical mass measurement before LIGO's detection were about a dozen of BHs in nearby X-ray binaries and their mass is well below 20 M$_\odot$ \cite{ozel2010,farr2011}.

Moreover, the first detections also provide us with a measurement of the  mass and spin of the  final (i.e. post-merger) BH and with some (still rather limited) information on the spins of the merging BHs. Finally, the estimate of the redshift of merging BHs represents the first step towards reconstructing the cosmic BH merger rate. Table~\ref{tab:table1} is a summary of some of the main quantities obtained from LIGO-Virgo observations with their 90\% credible intervals (according to \cite{abbott2016c,abbott2017a,abbott2017b,abbott2017c}). In the next sections, we will review our state-of-the-art knowledge of BH astrophysics and we will discuss the implications of the first five detections on the mass spectrum of BHs and on the formation channels of merging BHs.

\section{Implications for the mass spectrum of black holes}
Three of the first four detected events (GW150914, GW170104, and GW170814) involve BHs with mass $m_{\rm BH}>30$ M$_\odot$. If these BHs have stellar origin, such large masses can be understood only in terms of quenched stellar winds and failed supernova (SN) explosions. 

\subsection{Supernovae}
Our understanding of core-collapse SNe is still far from satisfactory. The outcome of a core-collapse SN depends on the mechanism of neutrino transfer, as well as on the impact of stellar rotation and magnetic fields \cite{foglizzo2015}. Moreover, two- and three-dimensional hydrodynamical simulations are still too computationally expensive to be used for a systematic study of the remnant mass (e.g. \cite{ertl2016}). Despite these issues, recent models of core-collapse SNe have started converging on some common ground. 

The models of Fryer et al. (2012, \cite{fryer2012}) suggest that the final mass of a BH depends mainly on the mass of the Carbon-Oxygen (CO) core of its progenitor star and on the amount of fallback. For CO core masses $m_{\rm CO}\gtrsim{}8$ M$_\odot$ a star is expected to directly collapse to a BH, without SN explosion. In this case, the entire final mass of the star collapses to a BH, producing a more massive BH than in the case of a SN explosion. 

Several authors suggest that the CO mass is not sufficient to understand the final fate of a star. For example, \cite{oconnor2011,ugliano2012} introduce the compactness parameter, defined as
\begin{equation}
\xi{}_M=\frac{M/{\rm M}_\odot}{R(M)/1000\,{}{\rm km}},
\end{equation}
where $M$ is a reference mass (usually $M=2.5$ M$_\odot$) and $R(M)$ is the radius which encloses this mass. The compactness parameter is usually measured at the onset of collapse \cite{ugliano2012}. A large value of $\xi{}_M$ ($\xi{}_{2.5}\gtrsim{}0.2$, according to \cite{horiuchi2014}) indicates that the star is going to collapse to a BH directly. 

Limongi (2017, \cite{limongi2017}) has shown that there is a strong correlation between the final CO mass and the value of $\xi{}_{2.5}$ (see figure~20 of \cite{limongi2017}), suggesting that models based on the CO mass and on the compactness parameter are in reasonable agreement. The main difference is that models based on the compactness parameter \cite{oconnor2011,ugliano2012} and even more complex models (e.g. the two parameter criterion by \cite{ertl2016}) predict islands of ``explodability''\footnote{This means that even very massive stars might have $\xi{}_{2.5}<0.2$ and thus are expected to explode as SNe, or vice versa even rather light stars might have $\xi{}_{2.5}>0.2$ and collapse directly.}, while models based on the CO mass \cite{fryer2012} predict a rather smooth transition between exploding and non-exploding models (as a function of the final mass of the star).

These differences must not be neglected, but overall we can predict a general trend:
\begin{itemize}
\item{}The details of the SN mechanism and the efficiency of fallback are particularly important for stars with mass $<<30$ M$_\odot$, which are not particularly affected by stellar winds.
\item{}For massive ($>30$ M$_\odot$) stars, stellar winds are extremely important and determine the final fate of a star.
\end{itemize}

In addition, pulsational-pair instability and pair instability SNe play a crucial role for very massive metal poor stars \cite{woosley2017}: 
\begin{itemize}
\item{}Stars with $30\lesssim{}M_{\rm He}/{\rm M}_{\odot}\lesssim{}60$  (where $M_{\rm He}$ is the mass of the Helium core), are expected to undergo pulsational pair instability, losing a large fraction of their mass.
\item{}If $60\lesssim{}M_{\rm He}/{\rm M}_{\odot}\lesssim{}130$, a star is expected to explode as pair-instability SN, leaving no remnant.
\item{}For $M_{\rm He}\gtrsim{}130$ M$_\odot$, the star is so massive that the collapse induced by pair production cannot be reversed: the star is expected to directly collapse to a very massive BH.
\end{itemize} 

\begin{figure}
\begin{center}
\includegraphics[width=14cm]{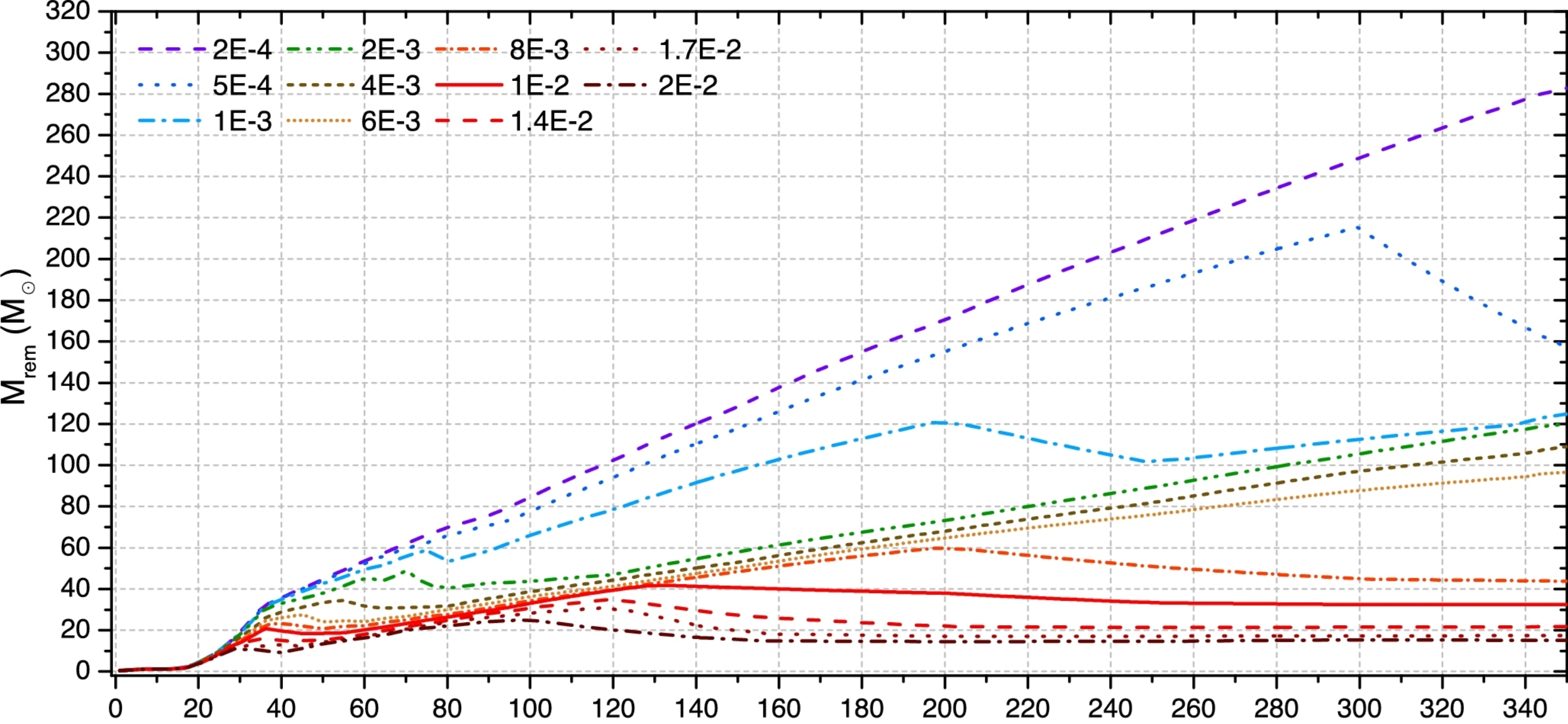}
\includegraphics[width=14.2cm]{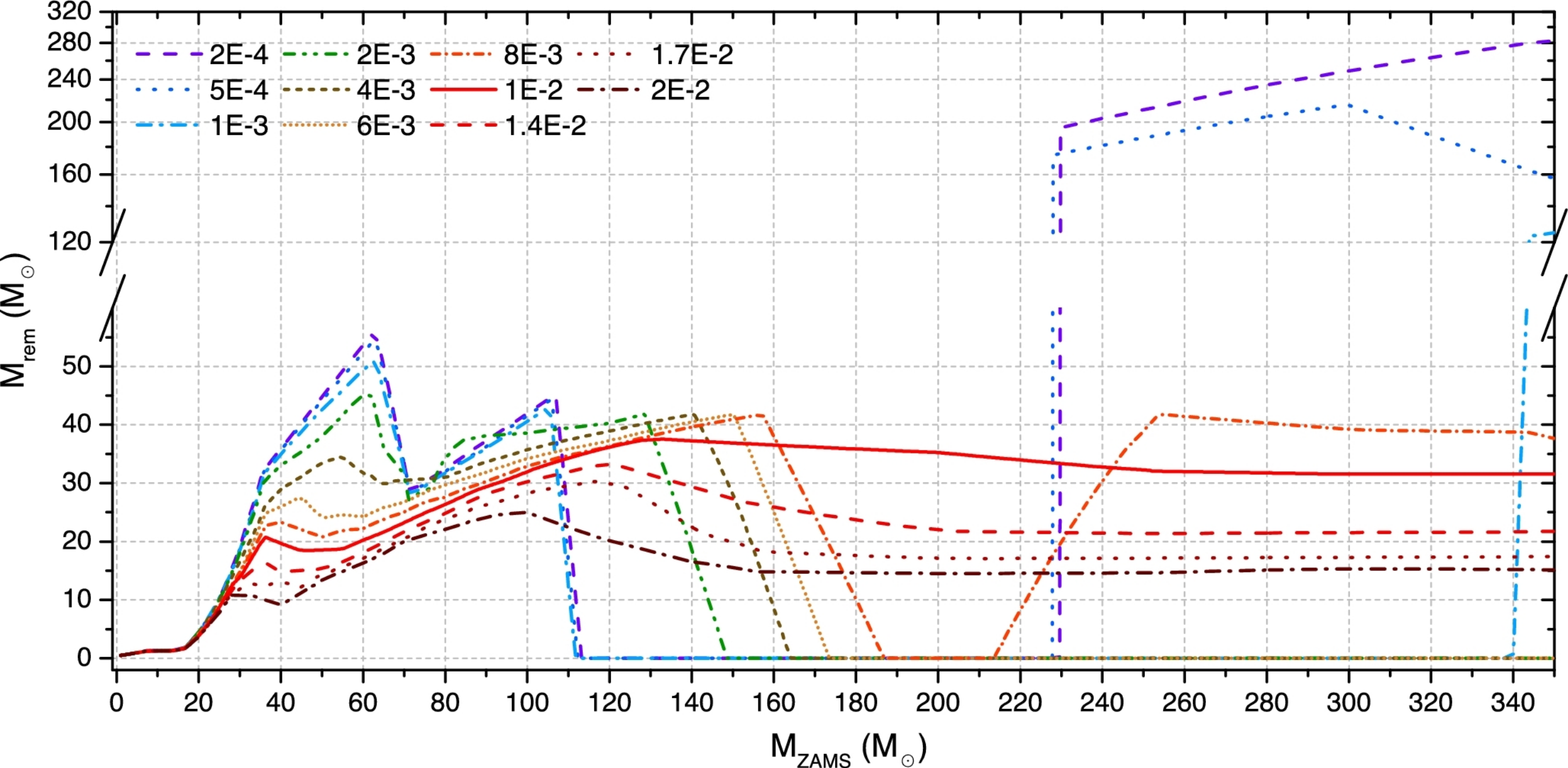}
\end{center}
\caption{\label{fig:spera}Mass of the compact object as a function of the zero age main sequence (ZAMS) mass of the progenitor star for 11 different metallicities (from $Z=2\times{}10^{-4}$ to $Z=0.02$). Upper (lower) panel: pair instability and pulsational pair instability SNe are not (are) included in the model. Adapted from Figures~1 and 2 of \cite{spera2017}.}
\end{figure}

Figure~\ref{fig:spera} clearly shows the impact of pair instability and pulsational pair instability SNe on the final remnant mass.

%%%%%%%%%%%%%%%%%%%%%%%%%%%%%%%%%%%%%%%%%%%%%%%%%%%%%%%TABLE%%%%%%%%%%%%%%%%%%%%%%%%%%%%%%%%%%%%%%%%%%%%%%%%%%%%%%%%%%%%%%%%%%%%%
\begin{table}
\caption{\label{tab:table1}Main observed properties of published GW detections (BH mergers only).}
\begin{center}
\begin{tabular}{lllllll}
\br
                        & GW150914             & LVT151012           & GW151226                & GW170104               &  GW170608       & GW170814\\
\mr
$m_1$ (M$_\odot$)         & $36.2^{+5.2}_{-3.8}$    & $23^{+18}_{-6}$       & $14.2^{+8.3}_{-3.7}$   &   $31.2^{+8.4}_{-6.0}$  & $12^{+7}_{-2}$   & $30.5^{+5.7}_{-3.0}$ \vspace{0.1cm}\\
$m_2$ (M$_\odot$)         & $29.1^{+3.7}_{-4.4}$    & $13^{+4}_{-5}$        & $7.5^{+2.3}_{-2.3}$    &    $19.4^{+5.3}_{-5.9}$ & $7^{+2}_{-2}$    & $25.3^{+2.8}_{-4.2}$ \vspace{0.1cm}\\
$m_{\rm chirp}$ (M$_\odot$)  & $28.1^{+1.8}_{-1.5}$    & $15.1^{+1.4}_{-1.1}$  & $8.9^{+0.3}_{-0.3}$    &   $21.1^{+2.4}_{-2.7}$  & $7.9^{+0.2}_{-0.2}$ & $24.1^{+1.4}_{-1.1}$ \vspace{0.1cm}\\
$m_{\rm TOT}$ (M$_\odot$)           & $65.3^{+4.1}_{-3.4}$    & $37^{+13}_{-4}$       & $21.8^{+5.9}_{-1.7}$   &    $50.7^{+5.9}_{-5.0}$ & $19^{+5}_{-1}$ & $55.9^{+3.4}_{-2.7}$ \vspace{0.1cm}\\
$m_{\rm fin}$ (M$_\odot$)   & $62.3^{+3.7}_{-3.1}$    & $35^{+14}_{-4}$       & $20.8^{+6.1}_{-1.7}$   &    $48.7^{+5.7}_{-4.6}$ & $18.0^{+4.8}_{-0.9}$ & $53.2^{+3.2}_{-2.5}$ \vspace{0.3cm}\\
$\chi{}_{\rm eff}$         & $-0.06^{+0.14}_{-0.14}$ & $0.0^{+0.3}_{-0.2}$    & $0.21^{+0.20}_{-0.10}$ &  $-0.12^{+0.21}_{-0.30}$ & $0.07^{+0.23}_{-0.09}$ & $0.06^{+0.12}_{-0.12}$ \vspace{0.1cm}\\
$S_{\rm f}$               & $0.68^{+0.05}_{-0.06}$   & $0.66^{+0.09}_{-0.10}$ & $0.74^{+0.06}_{-0.06}$ &  $0.64^{+0.09}_{-0.20}$  & $0.69^{+0.04}_{-0.05}$ & $0.70^{+0.07}_{-0.05}$ \vspace{0.3cm}\\
$d_{\rm L}$ (Mpc)         & $420^{+150}_{-180}$     & $1000^{+500}_{-500}$   & $440^{+180}_{-190}$    &  $880^{+450}_{-390}$    & $340^{+140}_{-140}$ & $540^{+130}_{-210}$ \vspace{0.1cm}\\
$z$                     & $0.09^{+0.03}_{-0.04}$   & $0.20^{+0.09}_{-0.09}$ & $0.09^{+0.03}_{-0.04}$  &   $0.18^{+0.08}_{-0.07}$ & $0.07^{+0.03}_{-0.03}$ & $0.11^{+0.03}_{-0.04}$\vspace{0.1cm}\\
\br
\end{tabular}
\end{center}
\footnotesize{$m_1$ ($m_2$): mass of the primary (secondary) BH; $m_{\rm chirp}$ ($m_{\rm TOT}$): chirp (total) mass of the binary; $m_{\rm fin}$: mass of the final BH; $\chi_{\rm eff}$: effective spin; $S_{\rm fin}$: spin of the final BH; $d_{\rm L}$: luminosity distance; $z$: redshift. From left to right: GW150914, LVT151012, GW151226, GW170608, GW170104, GW170814. For all properties, we report median values with 90\%{} credible intervals (\cite{abbott2016c,abbott2017a,abbott2017b,abbott2017c}). Source-frame masses are quoted.} 
\end{table}
%%%%%%%%%%%%%%%%%%%%%%%%%%%%%%%%%%%%%%%%%%%%%%%%%%%%%%%%%%%%%%%%%%%%%%%%%%%%%%%%%%%%%%%%%%%%%%%%%%%%%%%%%%%%%%%%%%%%%%%%%%%%%%%

%%%%%%%%%%%%%%%%%%%%%%%%%%%%%%%FIGURE%%%%%%%%%%%%%%%%%%%%%%%%%%%%%%%%%%
%%%%%%%%%%%%%%%%%%%%%%%%%%%%%%%%%%%%%%%%%%%%%%%%%%%%%%%%%%%%%%%%%%%%%%%

\subsection{Stellar winds}
Mass loss by stellar winds is important because it determines the final mass of a star: according to the core-collapse SN models we  discussed in the previous section, stars with a larger final mass ($\gtrsim{}30$ M$_\odot$, \cite{fryer1999,mapelli2009}) are more likely to undergo a failed SN, leaving a larger remnant.

In the last decade, it has become increasingly clear that stellar winds depend on metallicity, not only during the main sequence (MS) but also afterwards \cite{vink2001,vinkdekoter2005,graefener2008,vink2011}. Another important ingredient is the Eddington factor $\Gamma=L_\ast/L_{\rm Edd}$ (where $L_{\ast}$ and $L_{\rm Edd}$ are the luminosity of the star and its Eddington limit, respectively). Most population-synthesis models account for the metallicity dependence of stellar winds, but neglect the impact of $\Gamma$, with few exceptions (e.g. \cite{spera2015,spera2017,mapelli2017,mapelli2018,giacobbo2018,giacobbo2018b}). According to Chen et al. (2015, \cite{chen2015}), mass loss by stellar winds can be modelled as $\dot{M}\propto{}Z^\alpha$, with
\begin{equation}\label{eq:mdot}
\alpha=\left\{
\begin{array}{ll}
0.85                     & \quad{}\quad{}\quad{}{\rm if\quad{}} \Gamma<2/3\\ 
2.45-2.40\,{}\Gamma{}     & \quad{}\quad{}\quad{}{\rm if\quad{}}1\geq{}\Gamma{}\geq{}2/3\\
0.05                      & \quad{}\quad{}\quad{}{\rm if\quad{}}\Gamma>1
\end{array} 
\right.
\end{equation}

This indicates that the dependence of $\dot{M}$ on metallicity becomes negligible if a star is radiation pressure dominated (i.e. $\Gamma{}$ approaches 1).
 
Figure~\ref{fig:spera} shows the mass spectrum of BHs obtained with the mass-loss prescription given in equation~\ref{eq:mdot} plus the delayed core-collapse SN model explained in \cite{fryer2012}. According to this model, the lower the metallicity is, the larger the maximum possible mass of the compact remnant. %Combined with a core-collapse SN model \cite{oconnor2011,fryer2012,ugliano2012,ertl2016}, this implies that the lower the metallicity is, the larger

Recent studies \cite{petit2017} show that a surface magnetic field has a similar effects to metallicity, quenching mass loss. Also, rotation cannot be neglected: on the one hand, stellar  winds are enhanced in fast rotating massive stars, on the other hand chemical mixing induces the formation of a larger Helium (and Carbon-Oxygen) core \cite{limongi2017}. While stronger stellar winds should translate into smaller remnant masses (see previous subsection), larger core masses might imply larger remnant masses. To complicate the situation even more, pair instability and pulsational pair instability SNe are expected to occur at smaller zero age MS (ZAMS) masses if the Helium core mass is larger, possibly reducing the number and mass of BHs \cite{limongi2017}. %the minimum zero age MS (ZAMS) mass for which a star undergoes a pair instability or a pulsational pair instability SN is smaller if the Helium core mass is larger, possibly reducing the number and mass of BHs \cite{limongi2017}.

\subsection{Primordial BHs}
Primordial BHs (e.g. \cite{bird2016,carr2016, inomata2016})  are predicted to form from gravitational instabilities in the very early Universe ($<1$ s after the Big Bang), via several mechanisms. The mass of a primordial BH is approximately connected with the horizon mass at the time of formation $\sim{}c^3\,{}t/G{}$, where $c$ is the speed of light, $G$ is the gravitational constant and $t$ is the time elapsed from the Big Bang \cite{carr2016}. According to this relation, BHs with mass $\sim{}30$ M$_\odot$ could have formed at $t\sim{}1.5\times{}10^{-4}$ s.
%their allowed mass range is $\sim{}1-1000$ M$_\odot$ \cite{carr2016}\footnote{Considering only BHs which produce gravitational waves in the LIGO-Virgo frequency band when they merge}. Thus, they might have a mass consistent with the observed gravitational wave events.

\subsection{Take home message for BH masses}
In summary, our knowledge of the BH mass spectrum is still hampered by large uncertainties (on stellar winds, on SNe, and on the existence of primordial BHs). However, we can conclude that the progenitors of GW150914, GW170104 and GW1708114 were relatively metal-poor massive stars (e.g. \cite{abbott2016d,belczynski2016, mapelli2016, spera2017}), unless we invoke primordial BHs or BHs born from previous mergers \cite{gerosaberti2017}.

\section{Field or dynamical origin?}
The mechanisms driving the formation of close BH binaries are still matter of intense debate. Two main channels have been proposed: the evolution of isolated BH binaries and the dynamical formation in star clusters. 

\subsection{Isolated binary evolution}
According to the isolated binary evolution scenario, two massive stars which form gravitationally bound evolve through several binary evolution processes, till they become BHs and (if they are sufficiently close) they merge. Unfortunately, we have a rather poor knowledge of several binary evolution processes, especially mass transfer and common envelope \cite{ivanova2013}. Two stars with a ZAMS mass $\gtrsim{}30$ M$_\odot$ are expected to reach photospheric radii of several thousand solar radii when they become super-giant stars. If the initial semi-major axis of a binary is of the same order of magnitude as the maximum stellar radii, the binary is expected to prematurely merge, or to transfer mass through Roche-lobe overflow (if one of the two stars fills its Roche lobe), or to undergo a common envelope phase (e.g. if both stars fill their Roche lobes). During common envelope, the cores of the two stars are engulfed in the same envelope and are expected to spiral in, because of gas drag. If nothing removes the envelope, the two cores keep spiralling in till they merge: the binary produces only a single BH. In contrast, if the common envelope is ejected (e.g. by conversion of the orbital energy of the cores into thermal energy), the final binary is a naked binary (without envelope) and has a very short final semi-major axis. If SN kicks do not disrupt it, the naked binary might evolve into a double BH binary, sufficiently close to merge within a Hubble time. In contrast, binaries with much larger initial semi-major axes remain `detached binaries', i.e. they avoid mass transfer and common envelope, but are too loose to produce merging BHs. 

Models used in population-synthesis codes do not capture the complex physics of common envelope. The most used formalism depends on a free parameter ($\alpha{}$, describing the efficiency of conversion of orbital energy into thermal energy \cite{webbink1984}), which is uncertain to a factor of $\gtrsim{}10$.

Alternative models \cite{marchant2016,demink2016,mandel2016} predict that massive fast rotating metal-poor stars might produce merging BH binaries even without common envelope evolution. In fact, fast rotating stars might remain almost chemically homogeneous during their evolution. As a consequence, their stellar radii remain significantly smaller than in the non-rotating case. The chemically homogeneous evolution model can account only for the formation of nearly equal-mass merging BH binaries, with large mass (BH masses $>20$ M$_\odot$) and aligned spins.

In contrast, the common-envelope model (e.g. \cite{dominik2012,belczynski2016,stevenson2017,mapelli2017,giacobbo2018}) predicts the formation of merging BH binaries with both small and large masses (BH masses $>3$ M$_\odot$), accounting for both GW150914-like systems and GW151226-like systems. Strongly unequal-mass systems ($m_2/m_1\ll{}0.5$) are quite unlikely even in the common-envelope model. Common-envelope and tides are expected to align the spins of the two stars with the orbital angular momentum of the binary, although SN explosions or other mechanisms might misalign the final BH spins (e.g. \cite{oshaughnessy2017,wysocki2017}).

\subsection{Dynamical formation}
According to the dynamical formation scenario, two BHs can enter the same binary through a dynamical mechanism (for example, a dynamical exchange \cite{hills1980}). Dynamical mechanisms (with the exception of Kozai-Lidov resonance, \cite{kozai1962,lidov1962}) are efficient only if stars are in dense environments, such as young star clusters, open clusters, globular clusters and nuclear star clusters. On the other hand, massive stars (which are BH progenitors) form preferentially in dense environments \cite{portegieszwart2010}, suggesting that dynamical mechanisms are crucial for the formation and evolution of BH binaries.

Three-body encounters (i.e. close encounters between a binary system and an intruder star) affect the orbital properties of binaries, such as semi-major axis and eccentricity. Hard binaries (i.e. binaries with binding energy larger than the average kinetic energy of intruders) are expected to {\it harden} as a consequence of three-body encounters, meaning that their semi-major axis shrinks \cite{heggie1975}. As a consequence of dynamical hardening, a BH binary might become sufficiently close as to merge by gravitational wave emission within a Hubble time. Thus, dynamics can boost the merger rate.

{\it Exchanges} are three-body encounters during which the intruder replaces one of the former members of the binary. The probability of an exchange is particularly high if the mass of the intruder is larger than the mass of one of the members of the binary \cite{hills1980}. Thus, BHs and BH progenitors, being among the most massive objects in star clusters, are particularly effective in capturing companions through exchanges. Ziosi et al. (2014, \cite{ziosi2014}) find that $\gtrsim{}90$ per cent of double BH binaries in young star clusters and open clusters form through dynamical exchanges.

Three-body encounters and exchanges have a crucial impact on a BH binary population: they tend to produce more massive binaries, with non-zero eccentricity, and to randomize their spins \cite{ziosi2014,rodriguez2015,rodriguez2016,mapelli2016,banerjee2017a,banerjee2017b,chatterjee2017}. A nearly isotropic distribution of BH spins is expected from dynamics. 

In young massive metal-poor clusters, dynamics can even lead to the formation of intermediate-mass BHs (i.e. BHs with mass $10^2\le{}m_{\rm BH}/{\rm M}_\odot\le{}10^4$), through the {\it runaway collision mechanism} \cite{portegieszwart2004,giersz2015,mapelli2016}.

Unlike the previous processes, the Kozai-Lidov (KL) resonance \cite{kozai1962,lidov1962} might be important also in low-density environments. A hierarchical triple undergoes KL resonance if the inclination between the orbital plane of the inner binary and that of the outer body is non zero. Perturbations by the outer body induce periodic oscillations of both the eccentricity of the inner binary and the inclination between the two orbital planes, without changing the semi-major axis. The increase of eccentricity might speed up the merger of the inner binary and might even produce systems with non-zero eccentricity in the LIGO-Virgo frequency range \cite{antonini2012,kimpson2016,antonini2017}. KL resonances are deemed to be a quite common process, since the occurrence of triple systems in massive stars might be as high as $\sim{}50$ per cent (see \cite{toonen2017} and references therein). In addition, \cite{antonini2012}  propose that KL oscillations might also occur in very exotic triple systems, composed of an inner BH binary and (as a tertiary body) a super-massive BH, in nuclear star clusters.

In summary, dynamics has multiple effects on merging BHs: (i) dynamics tends to produce more massive BH binaries, or even intermediate-mass BH binaries; (ii) dynamically formed binaries might have non zero eccentricity, even close to the merger; (iii) we expect a nearly isotropic distribution of spins in dynamically formed binaries; (iv) dynamics should boost the merger rate (e.g. through hardening).

Unfortunately, our knowledge of the dynamical formation scenario is hampered by several issues. While many papers have been published about the dynamics of BHs in old globular clusters (e.g. \cite{sigurdsson1993,portegieszwart2000,downing2010,downing2011,tanikawa2013,rodriguez2015,rodriguez2016,zevin2017,askar2017}) only few papers deal with young star clusters (e.g. \cite{ziosi2014,mapelli2016,banerjee2017a,banerjee2017b}), which are the most common birthplace of massive stars, at least in the nearby Universe. The main reason is likely a numerical one: globular clusters are spherically symmetric, relaxed, gasless old systems which can be simulated with a fast Monte Carlo approach, while young star clusters are rather asymmetric, substructured, not yet relaxed, gas rich systems, requiring accurate and expensive direct N-body simulations to study them. %Moreover, there is no complete study of young star cluster dynamics in a cosmological context.
For these reasons, constraints on the dynamical formation scenario are still quite poor. %PARLARE DEI PROBLEMI DELLA DINAMICA

\subsection{Merger rates from dynamically formed versus isolated binaries}
\begin{figure}
\begin{center}
\includegraphics[width=14cm]{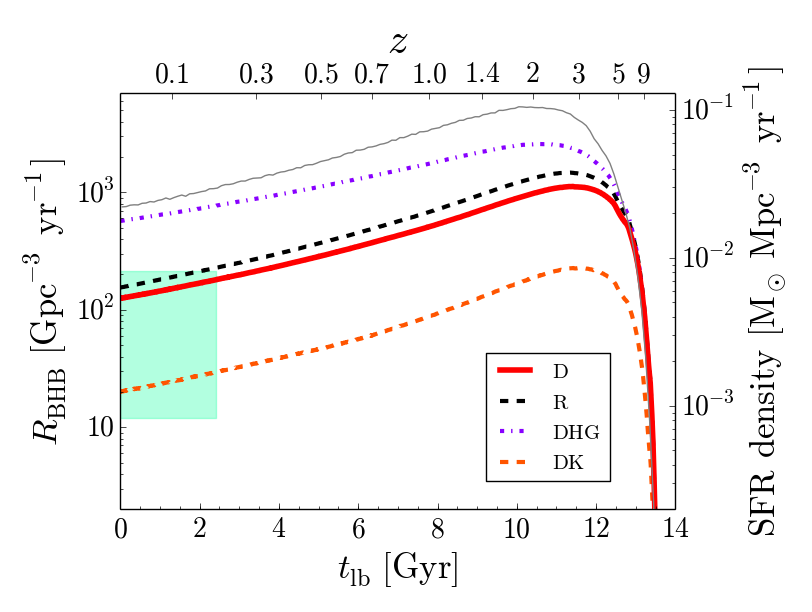}
\end{center}
\caption{\label{fig:cosmicrate} Cosmic merger rate density of BH binaries in the comoving frame ($R_{\rm BHB}$) as a function of the look-back time ($t_{\rm lb}$, bottom $x-$axis) and of the redshift ($z$, top $x-$axis) for several population synthesis models \cite{giacobbo2018} interfaced with the Illustris simulation \cite{vogelsberger2014}. Red solid line: model D (adopting a delayed SN model \cite{fryer2012}, common-envelope parameter $\alpha{}=1$, and fiducial distribution for the natal kicks of the BHs). Black dashed line: model R, same as model D but for a rapid SN model \cite{fryer2012}. Violet dot-dashed line: model DHG (same as model D but assuming that Hertzsprung gap donors can survive a common-envelope phase). Orange dashed line: model DK (same as model D but with maximum BH natal kicks). Green shaded area: local merger rate density inferred from the first LIGO-Virgo detections \cite{abbott2017a}. Right-hand $y-$axis: cosmic star formation rate density in the Illustris, shown as a thin grey line. 
 Adapted from Figure~1 of \cite{mapelli2017}.}
\end{figure}

The BH merger rate inferred from the first LIGO-Virgo detections is $\sim{}12-213$ Gpc$^{-3}$ yr$^{-1}$ \cite{abbott2017a}. Several studies report merger rates from their models, which can be compared to this estimate.
\begin{itemize}
\item[-]{\it Isolated binaries.} Local merger rates from isolated binaries span from $R_{\rm BHB}\sim{}10$ to $\sim{}800$ Gpc$^{-3}$ yr$^{-1}$ \cite{dominik2012,belczynski2016,mapelli2017,mapelli2018,giacobbo2018b}, depending on the assumptions about common-envelope efficiency, SN kicks and distribution of BH progenitors with redshift (see e.g. Figure~\ref{fig:cosmicrate}).
\item[-]{\it Globular clusters.} Merger rates from BH binaries in globular clusters are estimated to be  $R_{\rm BHB}\sim{}5$  Gpc$^{-3}$ yr$^{-1}$ \cite{rodriguez2016,askar2017,samsing2018}. This value is still uncertain but appears to be quite low with respect to the inferred rate. Actually, we expect the merger rate to be boosted in globular clusters by dynamical effects, but globular clusters represent a small fraction of the total stellar mass in galaxies (less than few per cent, \cite{harris2013}).
\item[-]{\it Young star clusters.} Merger rates from young star clusters and open clusters are extremely uncertain, because there are only few hundreds N-body simulations of these systems. The allowed range currently spans $\sim{}0.1-100$  Gpc$^{-3}$ yr$^{-1}$ \cite{ziosi2014,banerjee2017a,fujii2018}.
\item[-]{\it Nuclear star clusters.} The rate of mergers happening in nuclear star clusters was estimated to be $1-2$  Gpc$^{-3}$ yr$^{-1}$ by \cite{antonini2016}.  However, this number is also subject to large uncertainties. Most mergers in nuclear star clusters seem to involve relatively massive systems (similar to GW150914).
\end{itemize}

%%%%%%%%%%%%%%%%%%%%%%%%%%%%%%%FIGURE%%%%%%%%%%%%%%%%%%%%%%%%%%%%%%%%%%
%%%%%%%%%%%%%%%%%%%%%%%%%%%%%%%%%%%%%%%%%%%%%%%%%%%%%%%%%%%%%%%%%%%%%%%

\subsection{Take-home message for the formation channels}
From this Section, we have seen that masses and spins of merging BHs might help us distinguishing between the isolated binary evolution scenario and the dynamical formation scenario (e.g. from dynamics we expect more massive binaries with uniform spin distribution), but still too many issues and uncertainties affect theoretical models. New more sophisticated models are needed, especially for young star clusters. It is also important to apply model selection (see e.g. \cite{zevin2017}) between all existing models, in order to understand how many detections are needed to distinguish between proposed models.

\section{The cosmological context}
The first five published events have redshift spanning from $\sim{}0.07$ to $\sim{}0.2$. With the increase of sensitivity in the next observing runs, LIGO and Virgo will be able to observe BH mergers even at higher redshifts, while the third generation ground-based gravitational wave detectors (e.g. Einstein Telescope, \cite{punturo2010}) will enable us to reconstruct the entire merger history of BH binaries up to very high redshift ($z\sim{}8-10$).

A lot of work has been  done  recently to investigate the evolution of the BH merger rate with redshift (e.g. \cite{dominik2013,belczynski2016,lamberts2016,schneider2017,mapelli2017,mapelli2018}). Several studies suggest that the BH merger rate in the comoving frame is going to increase with redshift (Fig.~\ref{fig:cosmicrate}) with a trend depending on both the cosmic star-formation rate density and the metallicity evolution \cite{dominik2013,mapelli2017}.

This result is possibly very important for constraining BH formation channels. In fact, while the merger rate of stellar-born BHs approximately follows the cosmic star formation rate,  there is no reason for primordial BHs to behave in the same way (although the details depend on the delay time between the formation of the BHs and their merger). The key question here is how many detections we need to reconstruct the cosmic BH merger rate ``sufficiently well'' and what is the minimum sensitivity interferometers should reach to probe a large enough volume. Is design sensitivity of Advanced LIGO and Virgo enough to get some constraints from the merger history or should we wait for third generation ground-based detectors?\footnote{The paper by Fishbach et al. (2018, \cite{fishbach2018}), which was submitted after the first version of this proceeding was published, is the first quantitative attempt to address this question.}
\section{Conclusions}
The first five LIGO detections have shown that stellar BHs might be significantly more massive than we thought ($\sim{}30-40$ M$_\odot$). Recent models suggest that such massive BHs can arise only from the evolution of relatively metal-poor massive stars ($Z\lesssim{}0.3$ Z$_\odot$ \cite{mapelli2009,mapelli2010,belczynski2010,fryer2012,mapelli2013,spera2015}), or from gravitational instabilities in the early Universe (primordial BHs, \cite{carr2016}), or from previous mergers of stellar BHs \cite{gerosaberti2017}.

The cosmic merger rate of stellar-born BHs is expected to approximately scale with the cosmic star formation rate density \cite{dominik2013,mapelli2017}, while the merger rate of primordial BHs is expected to follow a different trend.  Future detections might be able to disentangle between stellar-born BHs and primordial BHs, if the cosmic BH merger rate history is reconstructed to a sufficiently high redshift (e.g. \cite{fishbach2018}).

Not only BH masses but also BH spins can give us possible clues to understand the formation channels of merging BHs (dynamical evolution model versus isolated binary evolution). However, the uncertainty on measured effective spins are still very large and models are also affected by several problems. For example, it is not clear how the magnitude of the spin of a BH depends on the magnitude of the spin of the progenitor star \cite{gerosaberti2017}. While it is known that dynamically formed binaries prefer misaligned spins, the spin distribution in isolated binaries is basically unconstrained. These uncertainties result in several degeneracies between models.

The era of gravitational-wave astrophysics has now begun, opening a completely new perspective for the study of BH demography. To meet this challenge, theoretical models must reach a better understanding of the astrophysical processes driving the formation of BH binaries. %, both for the dynamical and for the isolated formation channels.

%be significantly improved and updated, reaching a better understanding of the astrophysics 

%\begin{figure}[h]
%\begin{minipage}{14pc}
%\includegraphics[width=14pc]{spera17_fig1}
%\includegraphics[width=14pc]{spera17_fig2}
%\caption{\label{label}Figure caption for first of two sided figures.}
%\end{minipage}\hspace{2pc}%
%\end{figure}

%\begin{figure}[h]
%\includegraphics[width=14pc]{name.eps}\hspace{2pc}%
%\begin{minipage}[b]{14pc}\caption{\label{label}Figure caption for a narrow figure where the caption is put at the side of the figure.}
%\end{minipage}
%\end{figure}

\section*{Acknowledgments}
I thank the organizers and the participants of the 12$^{th}$ Edoardo Amaldi Conference on Gravitational Waves for the enlightening discussions. Numerical calculations have been performed through a CINECA-INFN agreement and through a CINECA-INAF agreement (Accordo Quadro INAF-CINECA 2017), providing access to resources on GALILEO and MARCONI at CINECA. MM  acknowledges financial support from the Italian Ministry of Education, University and Research (MIUR) through grant FIRB 2012 RBFR12PM1F, from INAF through grant PRIN-2014-14, and from the MERAC Foundation.
%\section*{Appendices}

\end{document}